\documentclass[twocolumn,english,prl,amssymb,aps,superscriptaddress,showpacs,twocolumn,amsmath,showkeys,floatfix]{revtex4-1}
\usepackage[T1]{fontenc}
\usepackage[latin9]{inputenc}
\usepackage{geometry}
\usepackage[active]{srcltx}
\usepackage{textcomp}
\usepackage{amsmath}
\usepackage{graphicx}
\usepackage{esint}

\makeatletter
\usepackage{babel}

\usepackage{babel}

\makeatother

\usepackage{babel}
\begin{document}

\title{Phase evolution of the direct detection noise figure of a non degenerate fiber phase sensitive amplifier}
\author{Tarek Labidi}\author{Ihsan  Fsaifes}
\affiliation{Laboratoire Aim\'e Cotton, Universit\'e Paris-Sud, ENS Paris-Saclay, CNRS, Universit\'e Paris-Saclay, 91405 Orsay, France}
\author{Weilin Xie}
\affiliation{Laboratoire Aim\'e Cotton, Universit\'e Paris-Sud, ENS Paris-Saclay, CNRS, Universit\'e Paris-Saclay, 91405 Orsay, France}
\affiliation{School of Optoelectronics, Beijing Institute of Technology, Beijing 100081, China}
\affiliation{State Key Laboratory of Advanced Optical Communication Systems and Networks, Shanghai Jiao Tong University, Shanghai, 200240, China}
\author{Debanuj Chatterjee}\author{Fabienne Goldfarb }
\affiliation{Laboratoire Aim\'e Cotton, Universit\'e Paris-Sud, ENS Paris-Saclay,
CNRS, Universit\'e Paris-Saclay, 91405 Orsay, France}
\author{Fabien Bretenaker}
\affiliation{Laboratoire Aim\'e Cotton, Universit\'e Paris-Sud, ENS Paris-Saclay,
CNRS, Universit\'e Paris-Saclay, 91405 Orsay, France}
\affiliation{Light and Matter Physics Group, Raman Research Institute, Bangalore 560080, India}

\begin{abstract} We experimentally investigate the evolution of the direct detection noise figure of a non degenerate phase sensitive amplifier based on nonlinear fiber, as a function of the relative phase between the signal, idler, and pump, all other parameters remaining fixed. The use of a fiber with a high stimulated Brillouin scattering threshold permits to investigate the full range of phase sensitive gain and noise figure without pump dithering. Good agreement is found with theory, both for signal only and combined signal and idler direct detections.
\end{abstract}


\maketitle
Optical amplifiers are at the heart of both digital optical telecommunication systems and analog microwave photonics links. However, the principles of quantum optics show that ordinary amplifiers, which amplify all the quadratures of a light mode with the same gain, lead to a degradation of the signal-to-noise ratio, due to the coupling of the amplified mode with vacuum modes \cite{Loudon2000}. This degradation, called the amplifier noise figure, cannot be smaller than 3 dB when the gain is large. 

Alternatively to such phase insensitive amplifiers (PIA), Caves \cite{Caves1982} has shown that a phase sensitive amplifier (PSA) can amplify the signal without degrading the signal-to-noise ratio. In this case, one quadrature of the considered mode undergoes a maximum gain $G_{\mathrm{max}}$ while the orthogonal quadrature is ``deamplified'' by a gain $G_{\mathrm{min}}=1/G_{\mathrm{max}}$. An incident quasi-classical state is then transformed into a Gaussian squeezed state, and the signal-to-noise ratio of any quadrature of the field is conserved through the amplifier, which thus exhibits a noise figure equal to 0 dB in homodyne detection. However, in a direct detection experiment, where both quadratures contribute to the signal, the noise figure of the amplifier depends on the phase of the signal that is amplified. It is minimum (resp. maximum) when the signal experiences the maximum (resp. minimum) gain. 

Such a dependence on the gain of the amplifier  noise figure in direct detection  has been observed in the first demonstrations of noiseless amplification in $\chi^{(2)}$ media \cite{Levenson1993a, Levenson1993b}. In these experiments, the noise figure was shown to vary from 7 dB to 1 dB when the gain of the quadrature was varied from -2 dB to 8 dB. 

More recently, there has been a growing interest in the development of PSA based on Kerr effect in $\chi^{(3)}$ media, and more precisely in optical fibers. This interest is driven by applications in communication technologies. Indeed, the availability of highly nonlinear fibers (HNLF) permits to build fiber parametric amplifiers exhibiting very large gains on a very broad bandwidth \cite{Tong2012,Tong2013,Karlsson2016}, in which all the interacting wavelengths -- pump(s), signal, and idler -- are in the telecommunication window and can thus propagate with very small losses in the single-mode fiber. This has led to the demonstration of in-line amplification with a noise figure close to 0 dB, much smaller than the 3 dB limit of PIA \cite{Lim2008,Tong2009}. However, such large gains require the use of powerful pumps and long fibers, leading to the occurrence of stimulated Brillouin scattering (SBS) that degrades the operation of the amplifier. The spectrum of the pump(s) is thus usually broadened by modulation with several tones in order to increase the Brillouin threshold \cite{Korotky1995,Hansryd2001}. The consequence of this technique is that the different frequency components of the modulated pump(s) do no longer have the same phase difference with the signal and idler, leading to a reduction of the maximum gain and an increase of the minimum gain \cite{Mussot2004}. Moreover, this has led to the difficulty to observe the dependence of the PSA noise figure on the relative phase between the interacting waves, contrary to the observation performed decades ago in $\chi^{(2)}$ crystals \cite{Levenson1993b}. In most fiber PSA literature, the relative phase of the waves is locked at the point of maximum gain and only the smallest value of the noise figure is measured \cite{Lim2008, Tong2010a,Tong2011a, Tong2011b, Malik2014}. In some papers \cite{Tong2010}, the changes in the noise figure are observed when the wavelength is varied, which also induces a change in the relative phase. Besides, some papers in the literature report measurements of the noise figure as a function of the input power or of the wavelength \cite{Tong2010, Malik2016}, but not as a function of the relative phase of the waves, all other parameters being fixed. Thus the aim of this Letter is to build a fiber-based PSA specially designed to allow us to scan the relative phase between the three interacting wave from $0$ to $2\pi$ and thus to explore the whole range of gains, from $G$ to $1/G$, while keeping all other parameters (wavelengths, pump power, signal and idler power, etc...) constant. We should thus be able to measure the evolution as a function of the relative phase of the corresponding noise figure evolution in a direct detection experiment.

\begin{figure}[htbp]
\centering
\includegraphics[width=0.8\linewidth]{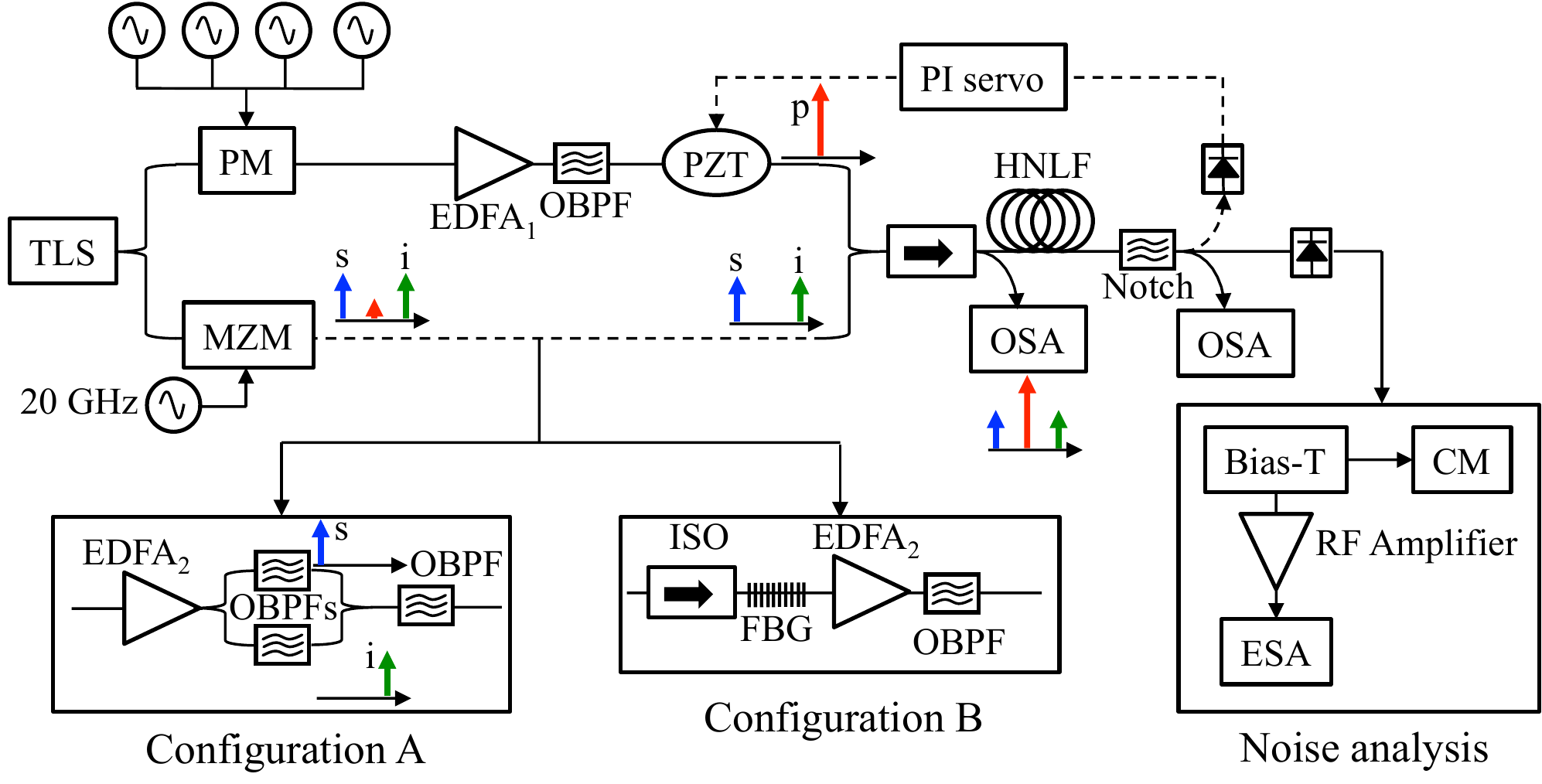}
\caption{Experimental setup. TLS: tunable laser source; EDFA: erbium-doped fiber amplifier, PZT: piezo-electric transducer; PM: phase modulator; MZM: Mach-Zehnder modulator; OBPF: optical bandpass filter; OSA: optical spectrum analyzer; FBG: fiber Bragg grating; HNLF: highly nonlinear fiber; ESA: electrical spectrum analyzer; CM: current meter.}
\label{Fig01}
\end{figure}

To achieve this aim, we built the pump degenerate parametric amplifier experiment sketched in Fig. \ref{Fig01}. A tunable semiconductor laser (Yenista Tunics) tuned at 1547\;nm is split into two arms by a 10\%/90\% coupler. The 10\% arm is used to create the degenerate pump and  is thus amplified by a 30-dBm-power EDFA and filtered by a 0.2-nm-bandwith bandpass filter to reduce amplified spontaneous emission noise. The 90\% arm is modulated by an intensity Mach-Zehnder modulator biased at minimum carrier transmission and driven at 20\;GHz. After filtering out the carrier, the two sidebands at $\pm20\;\mathrm{GHz}$ thus constitute the signal and the idler of the fiber-optic parametric amplifier, respectively. In a first configuration, the remaining carrier in this arm is filtered out by splitting the beam into two arms containing each a 50-pm-bandwidth bandpass filter (configuration A in Fig. \ref{Fig01}). To compensate for the losses associated with this filtering, a 20-dBm EDFA is inserted in the signal and idler arm of the experiment. The two arms are then recombined using another 10\%/90\% coupler and injected in the HNLF after an optical isolator. A notch filter is used at the output of the HNLF to remove the pump and observe the signal and/or idler spectra and power. All components are polarization-maintaining, ensuring that the three waves are co-polarized at the input of the HNLF.

In a first experiment, we use a standard 1011-m-long HNLF provided by OFS \cite{OFS}, with a nonlinear coefficient $\gamma=11.3\;(\mathrm{W.km})^{-1}$, a zero-dispersion wavelength $\lambda_{ZDW}=1547\;\mathrm{nm}$ and a dispersion slope $D^{\prime}=0.017\;\mathrm{ps.nm}^{-2}.\mathrm{km}^{-1}$. In order to increase the SBS threshold, we modulate the phase of the pump by inserting a 10 GHz bandwidth phase modulator in the pump arm . Once this modulator is driven at four frequencies (100, 300, 900 and 2700 MHz), we can inject a pump power $P_{\mathrm{p}}$ as large as 24 dBm in the HNLF without observing any SBS. We then inject equal signal and idler powers $P_{\mathrm{s}}=P_{\mathrm{i}}=-13\;\mathrm{dBm}$ into the fiber to probe the nonlinear amplification. In these conditions, we expect the phase sensitive gain to be given by \cite{Tong2010b}
\begin{equation}
G_{\mathrm{PSA}}=2G-1+2\sqrt{G(G-1)}\cos\Theta\ ,\label{Eq01}
\end{equation}
where $G$ is the PIA gain (when only the signal is launched in the fiber) and $\Theta=\phi_{\mathrm{s}}+\phi_{\mathrm{i}}-2\phi_{\mathrm{p}}$ is the relative phase between the signal, idler, and pump, whose phases are $\phi_{\mathrm{s}}$, $\phi_{\mathrm{i}}$, and $\phi_{\mathrm{p}}$, respectively. $G$ is given by
\begin{equation}
G=1+\left(1+\frac{\kappa^2}{4 g^2}\right)\sinh^2 gL\ ,\label{Eq01N1}
\end{equation}
where $\kappa=2\gamma P_{\mathrm{p}}-\Delta\beta$ is the total phase mismatch, $\Delta\beta$ the linear phase mismatch, and $g^2=(\gamma P_{\mathrm{p}})^2-(\kappa/2)^2$.

\begin{figure}[htbp]
\centering
\includegraphics[width=0.6\columnwidth]{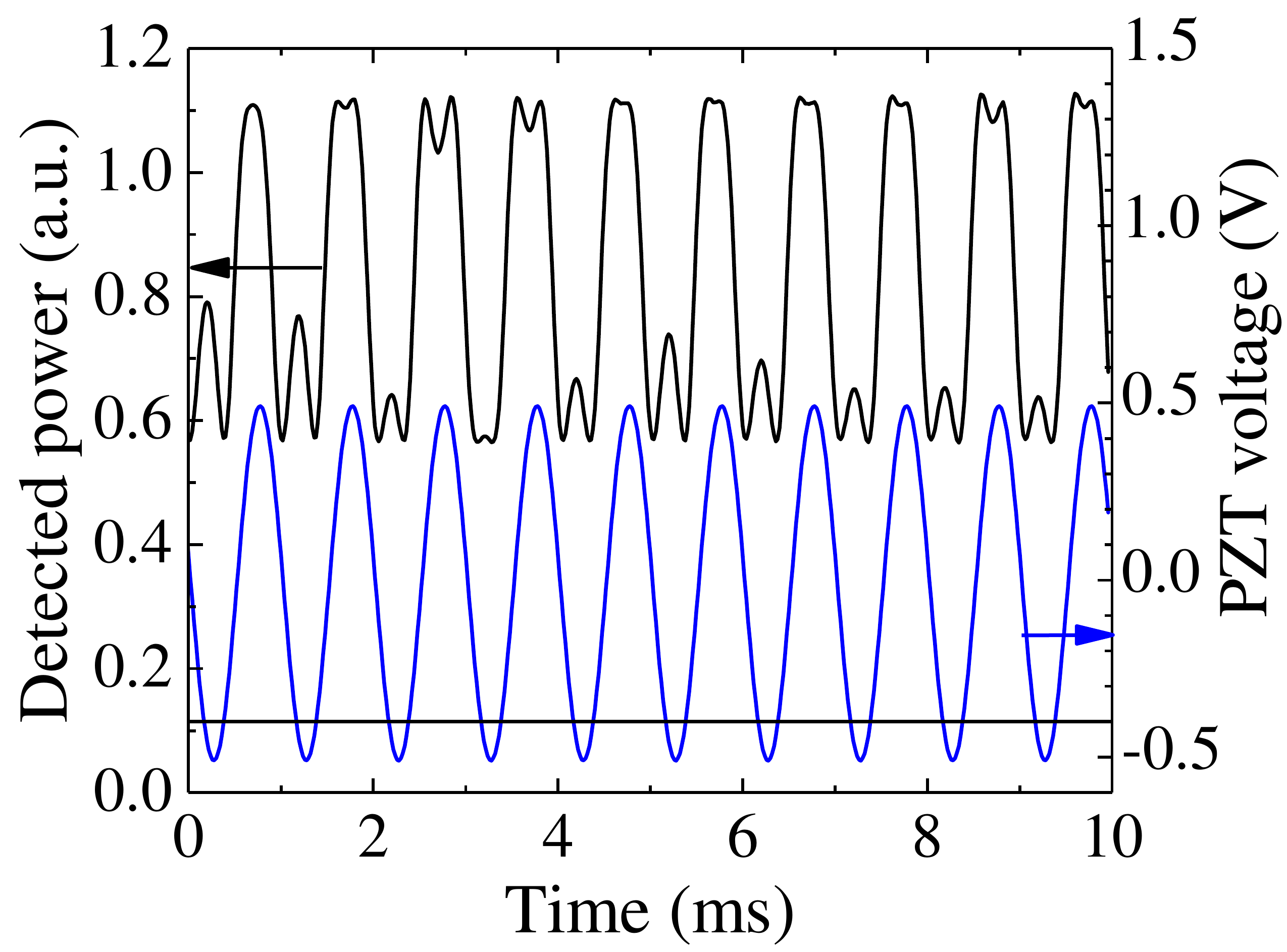}
\caption{Measured evolution of the total signal and idler amplified power versus time when the phase of the pump is sinusoidally modulated, in the presence of pump phase dithering. The horizontal line is the detected power when the pump is off. The bottom sinusoid is a picture of the voltage applied on the piezotransducer used to modulate the pump phase. The fiber is a $L=1011$-m-long standard HNLF fiber. $P_{\mathrm{p}}=24\;\mathrm{dBm}$. $P_{\mathrm{s}}=P_{\mathrm{i}}=-13\;\mathrm{dBm}$.}
\label{Fig02}
\end{figure}

From Eqs. (\ref{Eq01}) and (\ref{Eq01N1}) and with the parameters of our experiment, $G_{\mathrm{PSA}}$ should vary between $G_{\mathrm{max}}\simeq4 G=10\;\mathrm{dB}$ and $G_{\mathrm{min}}=1/G_{\mathrm{max}}=-10\;\mathrm{dB}$. Figure \ref{Fig02} shows a measurement of the sum of the signal and idler powers when we modulate the phase of the pump by applying a sinusoidal voltage to the piezoelectric transducer (PZT) shown in Fig.\;\ref{Fig01}, on which we have coiled a few meters of fiber in the pump arm. This figure shows that the gain is phase sensitive, as expected, but the maximum corresponds to a gain $G_{\mathrm{max}}=10\;\mathrm{dB}$ while the minimum to a gain $G_{\mathrm{min}}=7\;\mathrm{dB}$. Consequently, the amplifier does not exhibit the entire expected gain dynamics. In particular, the minimum gain is too large, compared to theoretically expected value, exhibiting amplification ($G_{\mathrm{min}}>1$) while it should correspond to deamplification ($G_{\mathrm{min}}<1$).

\begin{figure}[htbp]
\centering
\includegraphics[width=0.95\columnwidth]{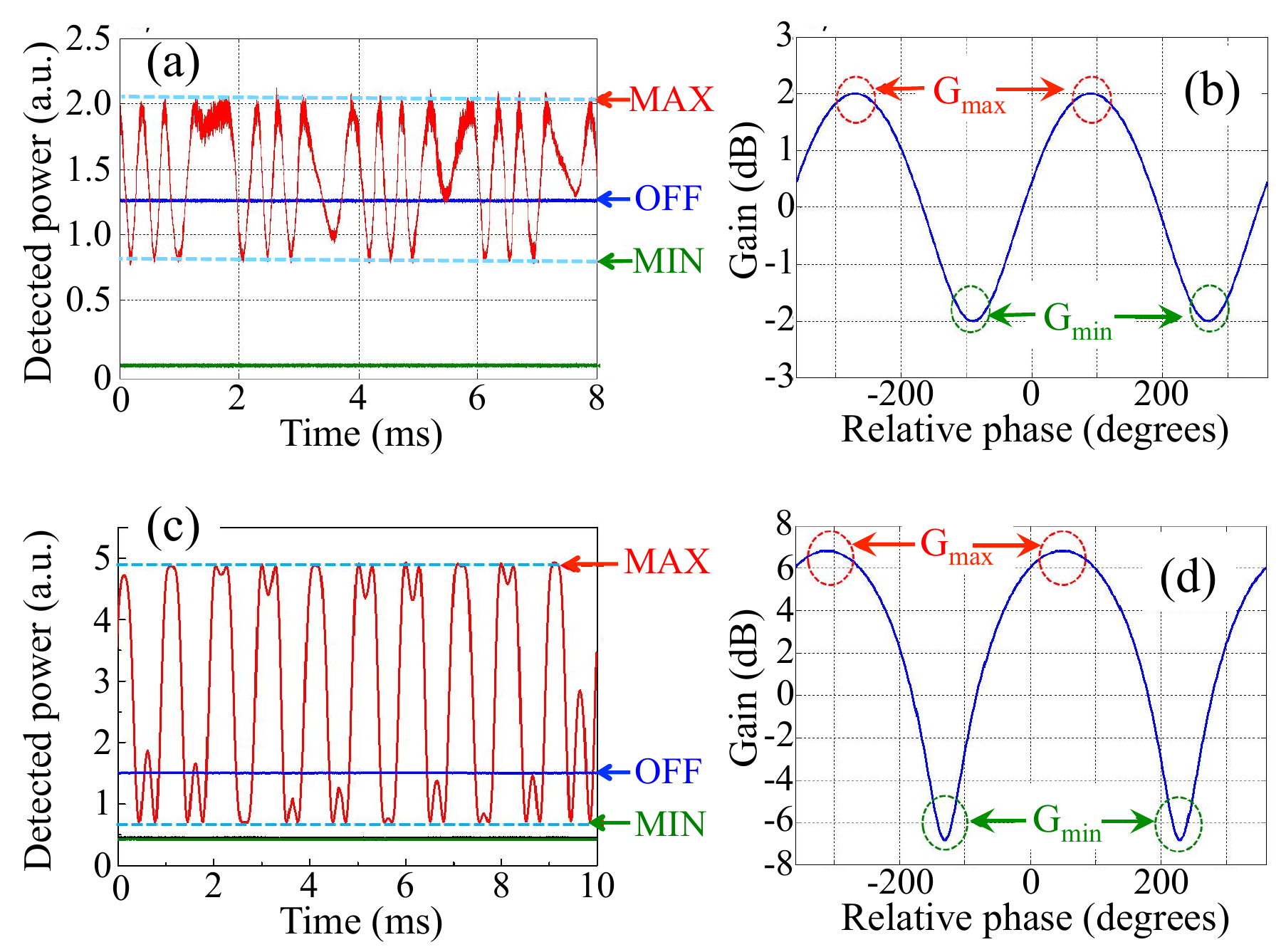}
\caption{(a,c) Measured and (b,d) calculated evolution of the total signal and idler power versus time when the pump phase is sinusoidally modulated, without pump phase dithering. (a,b) Standard HNLF with $L=200\;\mathrm{m}$, $P_{\mathrm{p}}=20\;\mathrm{dBm}$, and $P_{\mathrm{s}}=P_{\mathrm{i}}=-13\;\mathrm{dBm}$. (c,d) SPINE HNLF with $L=500\;\mathrm{m}$, $P_{\mathrm{p}}=24\;\mathrm{dBm}$, and $P_{\mathrm{s}}=P_{\mathrm{i}}=-3.5\;\mathrm{dBm}$. `OFF' represents the signal and idler power when the pump is off. In (a,c), the bottom horizontal line shows the residual pump power. The dashed lines correspond to the maximum and minimum gains. All powers are recorded in the same conditions, allowing to perform ratios to deduce the maximum and minimum gains.}
\label{Fig03}
\end{figure}

To check that this discrepancy with respect to theory is due to the fact that the pump contains many different tones, we reduce the fiber length down to $L=200\;\mathrm{m}$ and the  input  pump power down to $P_{\mathrm{p}}=20\;\mathrm{dBm}$ where we can switch off the pump phase dithering without exciting SBS. The corresponding evolution of the simultaneously detected signal and idler powers when the pump phase is sinusoidally modulated is reproduced in Fig.\;\ref{Fig03}(a). The values of the minimum and maximum gains can be extracted from this figure by comparing the levels labeled ``MAX'' and ``MIN'' with the blue line labeled ``OFF'', which corresponds to the power detected when the pump is switched off, and also talking into account the pump residual power that falls on the detector (bottom horizontal green full line). We measure a gain varying between $G_{\mathrm{min}}=-1.9\;\mathrm{dB}$ and $G_{\mathrm{max}}=2.0\;\mathrm{dB}$ when the relative phase $\Theta$ is scanned. The corresponding theoretical plot, in which $G_{\mathrm{min}}=-2.0\;\mathrm{dB}$ and $G_{\mathrm{max}}=2.0\;\mathrm{dB}$, is shown in Fig.\;\ref{Fig03}(b). The good agreement between the experimental and theoretical values of the minimum and maximum gains shows that the full range of gain values can be observed when one can avoid to broaden the pump spectrum by phase modulation.

In order to fully exploit the total gain dynamics of the amplifier  \cite{Lundstrom2013}, we thus shift to another fiber, also manufactured by OFS, which exhibits a larger SBS threshold \cite{OFS}. This so-called ``SPINE'' (Stable Phase Matching for Improved Nonlinear Efficiency) fiber has the following characteristics: $\gamma=8.7\;(\mathrm{W.km})^{-1}$, $\lambda_{ZDW}=1566\;\mathrm{nm}$ and $D^{\prime}=0.083\;\mathrm{ps.nm}^{-2}.\mathrm{km}^{-1}$. With a length $L=500\;\mathrm{m}$ and a launched pump power $P_{\mathrm{p}}=24\;\mathrm{dBm}$, no SBS was observed and we were able to observe the full PSA gain dynamics, from $G_{\mathrm{min}}=-6.0\;\mathrm{dB}$ to $G_{\mathrm{max}}=6.0\;\mathrm{dB}$, as shown in Fig.\;\ref{Fig03}(c). This is in good agreement with the theoretically expected values reproduced in  Fig.\;\ref{Fig03}(d), which also take the losses of the fiber into account. Moreover, the ``SPINE'' fiber exhibits a better longitudinal dispersion control, which guarantees stable phase matching conditions over the fiber length and making comparison between experiment and  theory  easier.

To stabilize the relative phase between the three waves, the lengths of the two arms of the experiment of Fig. 1 are balanced. The gain of the PSA can then be locked at almost any value between $G_{\mathrm{min}}$ and $G_{\mathrm{max}}$ by comparing the detected signal and idler power to a reference value. This reference value, and thus the locking point, are adjusted by sending the suitable correction signal to the piezoelectric transducer controlling the pump phase (see the dashed line in Fig.\;\ref{Fig01}). The PSA gain can thus be stabilized between its maximum and minimum values. This servo-loop is stable enough to perform gain and noise measurements, except very close to the maximum and minimum values of the gain where the error signal exhibits an extremum. Moreover, the signal and idler powers are continuously monitored during the noise measurements, allowing us to know exactly for which gain value the noise measurements are performed.

\begin{figure}[htbp]
\centering
\includegraphics[width=0.6\columnwidth]{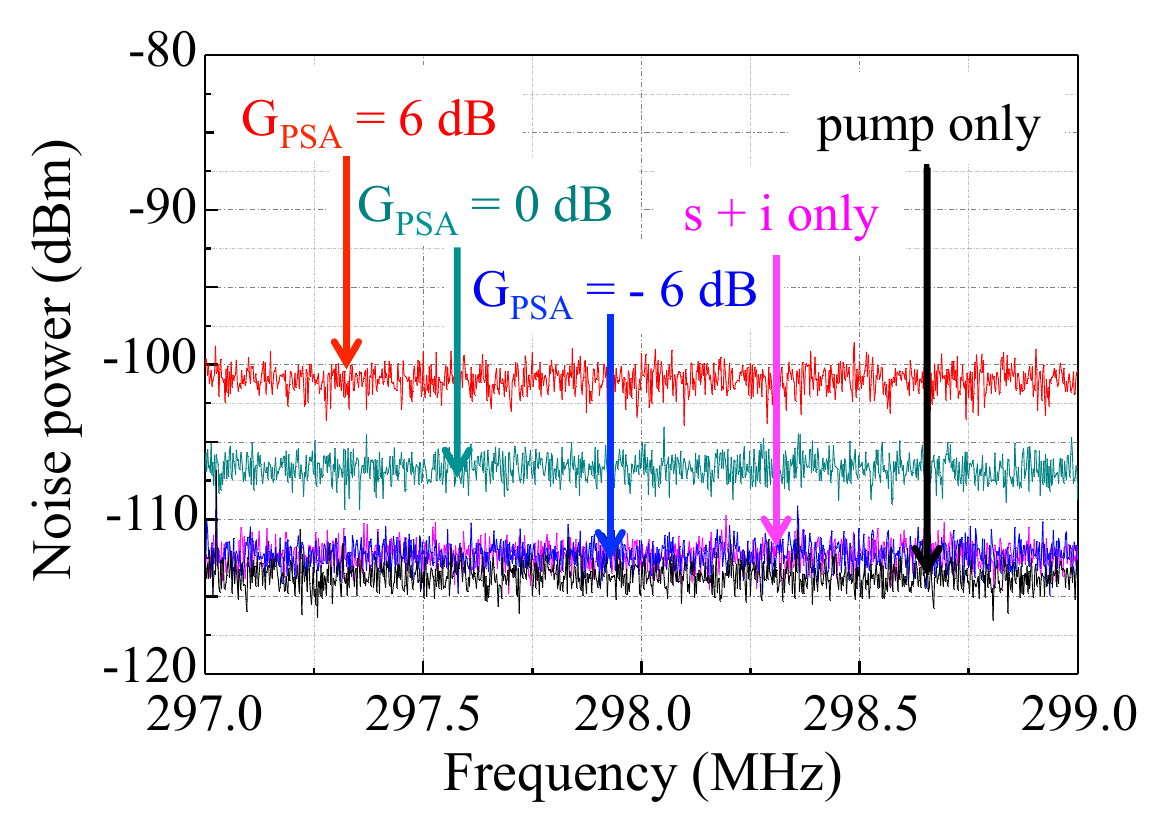}
\caption{Measured intensity noise spectra obtained in common signal and idler detection for different values of the phase sensitive gain,  with the signal and idler only (no pump), and in the presence of the pump only (no signal and idler). VBW = 10 Hz; RBW = 1 kHz.}
\label{Fig04}
\end{figure}

 By tuning the central wavelength and bandwidth of the notch filter located at the output of the HNLF we can reject the remaining pump and choose to detect either the power of the signal and the idler together (common detection) or the power of the signal alone. These amplified signal and/or idler powers are then detected by the noise analysis box of the setup of Fig.\,\ref{Fig01}. After a bias-T, the DC signal is  measured by a current meter while the noise component is amplified by a low noise RF-amplifier with 32 dB electrical gain and 500 MHz bandwidth. Finally, the noise levels are recorded by using an electrical spectrum analyzer (ESA) at 300 MHz Central frequency with 1 kHz resolution bandwidth and 10 Hz video bandwidth. It is worth noticing that a high efficiency photodetector as well as a low-noise RF-amplifier are important for a good measurement accuracy. Figure\;\ref{Fig04} shows examples of intensity noise spectra measured in common detection in such conditions. It is clear that the noise level depends on the gain experienced by the signal and the idler. 

The noise figure NF of the PSA can be evaluated from such measurements by comparing the output intensity noise level $N_{\mathrm{out}}$ with the input intensity noise level $N_{\mathrm{in}}$ through the relation, in logarithmic scale:
\begin{equation}
\mathrm{NF}(\mathrm{dB})=(N_{\mathrm{out}}(\mathrm{dB})-N_{\mathrm{in}}(\mathrm{dB}))-2G_{\mathrm{PSA}}(\mathrm{dB})\ ,\label{Eq02}
\end{equation}
where (dB) means that this calculation is valid when the quantities are expressed in dB. Moreover, one needs to be sure to be in the regime where the detected intensity noise is limited by the shot noise associated with the incident quasi-classical state. To this aim, with a variable attenuator, we checked the evolution of the detected noise power as a function of the generated photocurrent  when the PSA is switched off (pump off). In configuration A (see Fig.\;\ref{Fig01}), with an input signal and idler power $P_{\mathrm{s}}=P_{\mathrm{i}}=-8\;\mathrm{dBm}$ at the input of the nonlinear fiber, we could check that, when both signal and idler are detected, the noise level evolves linearly for photocurrents smaller 0.2\;mA. For larger photocurrents the noise level evolves quadratically with the photocurrent due to the predominance of the relative intensity noise induced by amplified spontaneous emission in the EDFA located on the signal and idler path of the experiment.  We have also checked that the noise of the residual pump after filtering is always smaller than the noise of the signal and idler. 
\begin{figure}[h]
\centering
\includegraphics[width=0.9\columnwidth]{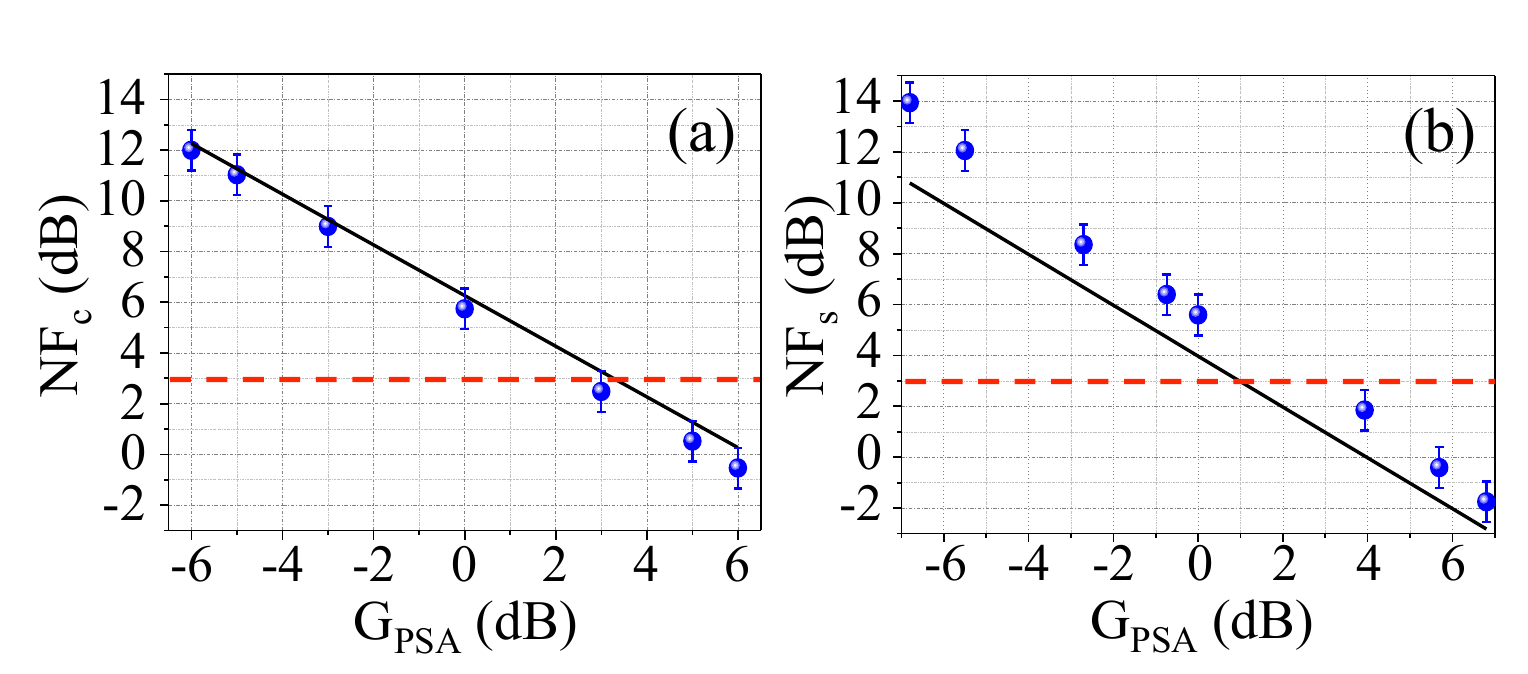}
\caption{Experimental (circles) and theoretical (solid line) evolutions of the PSA noise figure versus gain for (a) combined signal and idler direct detection and (b) signal direct detection. Powers at the input of the nonlinear fiber: (a) $P_{\mathrm{p}}=25\;\mathrm{dBm}$ and $P_{\mathrm{s}}=P_{\mathrm{i}}=-8\;\mathrm{dBm}$; (b) $P_{\mathrm{p}}=25.5\;\mathrm{dBm}$ and $P_{\mathrm{s}}=P_{\mathrm{i}}=-3.5\;\mathrm{dBm}$. Error bars:  ESA noise measurement uncertainty; Dashed lines: the classical 3 dB noise figure.}
\label{Fig05}
\end{figure}

In the case of combined signal and idler detection, the calculation of the evolution of the direct detection noise figure as a function of the PSA gain leads to the following expression \cite{Tong2012}, not taking the fiber losses into account:
\begin{equation}
\mathrm{NF}_{\mathrm{c,PSA}}=2\,\frac{2\,G-1}{G_{\mathrm{PSA}}}\ ,\label{Eq03}
\end{equation}
where $G_{\mathrm{PSA}}$ varies between $G_{\mathrm{min}}=(\sqrt{G}-\sqrt{G-1})^2$ and $G_{\mathrm{max}}=(\sqrt{G}+\sqrt{G-1})^2$ when $\Theta$ is varied. Equation (\ref{Eq03}) is plotted as a solid line in Fig.\;\ref{Fig05} for $G_{\mathrm{max}}=6\;\mathrm{dB}$, i. e. $G=1.9\;\mathrm{dB}$. The corresponding measured noise figures are plotted as circles in the same figure. They are obtained by applying Eq. (\ref{Eq02}) to noise levels measured at 300\;MHz, like the ones shown in Fig.\;\ref{Fig04}, for different values of the PSA gain. Very good agreement is observed between theory and experiment. In particular, for the maximum gain $G_{\mathrm{max}}=6\;\mathrm{dB}$, we obtain a noise figure equal to $-0.5\;\mathrm{dB}\pm0.8\;\mathrm{dB}$, which is significantly smaller than the value $\mathrm{NF}_{\mathrm{c,PIA}}=2-1/(2G-1)^2\simeq3\;\mathrm{dB}$ that would be obtained for a PIA with a 6 dB gain. On the contrary, for the minimum gain, the NF is close to $(2G-1)G_{\mathrm{max}}\simeq12\;\mathrm{dB}$, as expected from eq.\,(\ref{Eq03}).
 
In the case where only the signal is detected, we increased the signal and idler input powers by using configuration B in Fig.\;\ref{Fig01}. We could then reach signal and idler powers at the input of the nonlinear fiber equal to $P_{\mathrm{s}}=P_{\mathrm{i}}=-3.5\;\mathrm{dBm}$. The pump power is then equal to $P_{\mathrm{p}}=25.5\;\mathrm{dBm}$, leading to a maximum (resp. minimum) small signal PSA gain equal to $G_{\mathrm{max}}=6.8\;\mathrm{dB}$ (resp. $G_{\mathrm{min}}=-6.8\;\mathrm{dB}$), in agreement with the expected values. In this configuration, we could indeed reach higher signal and idler powers but unfortunately we could no longer completely filter out the excess noise due to the EDFA located in the signal/idler arm of the experiment, which was a few dB above shot noise. In spite of this, we performed noise figure measurements for different gain values. They are represented by circles in Fig.\;\ref{Fig05}(b), and compared with the following theoretical expression \cite{Tong2012}:
\begin{equation}
\mathrm{NF}_{\mathrm{s,PSA}}=\frac{2\,G-1}{G_{\mathrm{PSA}}}\ ,\label{Eq04}
\end{equation}
which leads to the solid line in Fig.\;\ref{Fig05}. One can see that the agreement remains qualitatively good, in spite of the super-Poissonnian noise introduced by the EDFA. In particular, for the maximum gain $G_{\mathrm{max}}=6.8\;\mathrm{dB}$, our measurement leads to $\mathrm{NF}_{\mathrm{s,PSA}}=-1.8\;\mathrm{dB}\pm0.8\;\mathrm{dB}$, which is well below the $3\;\mathrm{dB}$ limit expected in the case of a PIA. In this configuration, we could thus expect a noise figure equal to $1.2\;\mathrm{dB}$ if we detected the signal and idler together.

In conclusion, we have experimentally investigated the evolution of the noise figure of a PSA based on nonlinear optical fiber in direct detection, for both detection of the signal only and combined detection of the signal and the idler, as a function of the relative phase between the three waves, all other parameters (wavelengths, powers, etc.) being kept constant. The evolution of this noise figure with the relative phase, throughout the entire dynamics of the PSA gain, has been shown to be consistent with theory, and to be similar to the one observed in $\chi^{(2)}$ media. Moreover, these noise figures have been observed for incident signal and idler powers in the dBm range, compatible with analog microwave photonics applications where relatively large powers are mandatory. Besides, similar results should be obtainable for larger values of the gain if one uses longer SPINE fibers. Finally, the present results open interesting perspectives in cascading noiseless optical taps as demonstrated for $\chi^{(2)}$ PSAs \cite{Bencheikh1997}.

\end{document}